# High-temperature quantum kinetic effect in silicon nanosandwiches

N.T. Bagraev[1,2], V.Yu. Grigoryev[2], L.E. Klyachkin[1], A.M. Malyarenko[1], V.A. Mashkov[2], V.V. Romanov[2], and N.I. Rul'[2]

[1]*Ioffe Physical-Technical Institute, St. Petersburg 194021, Russia*

[2]*Peter the Great St. Petersburg Polytechnic University, St. Petersburg 195251, Russia*
E-mail: impurity.dipole@mail.ioffe.ru

The negative-$U$ impurity stripes confining the edge channels of semiconductor quantum wells are shown to allow the effective cooling inside in the process of the spin-dependent transport, with the reduction of the electron-electron interaction. The aforesaid promotes also the creation of composite bosons and fermions by the capture of single magnetic flux quanta on the edge channels under the conditions of low sheet density of carriers, thus opening new opportunities for the registration of the quantum kinetic phenomena in weak magnetic fields at high-temperatures up to the room temperature. As a certain version noted above we present the first findings of the high temperature de Haas-van Alphen, 300 K, quantum Hall, 77 K, effects as well as quantum conductance staircase in the silicon sandwich structure that represents the ultra-narrow, 2 nm, $p$-type quantum well (Si-QW) confined by the delta barriers heavily doped with boron on the $n$-type Si (100) surface.

## 1. Introduction

The Shubnikov–de Haas (ShdH) and de Haas–van Alphen (dHvA) effects as well as the quantum Hall effect (QHE) are quantum phenomena that manifest themselves at the macroscopic level and have attracted a lot of attention, because they reveal a deeper insight into the processes due to charge and spin correlations in low-dimensional systems [1,2]. Until recently the observation of these quantum effects in the device structures required ultra-low temperatures and ultra-high magnetic fields [3]. Otherwise there were the difficulties to provide small effective mass, $m^*$, and long momentum relaxation time, $\tau_m$, of charge carriers, which result from the so-called strong field assumption, $\mu B \gg 1$, where $\mu = e\tau_m/m^*$ is the mobility of the charge carriers. This severe criterion along with the condition $\hbar\omega_c \gg kT$ hindered the application of the ShdH-dHvA-QHE techniques to control the characteristics of the device structures in the interval between the liquid-nitrogen and room temperatures, where $\hbar\omega_c$ is the energy gap between adjacent Landau levels, $\omega_c = eB/m^*$ is the cyclotron frequency. Nevertheless, the ShdH oscillations were observed at room temperature in graphene, a single layer of carbon atoms tightly packed in a honeycomb crystal lattice, owing to the small effective mass of charge carriers, $\sim 10^{-4} m_0$, although the magnetic field as high as 29 T was necessary to be used because of relatively short momentum relaxation time [4–6]. Thus, the problem of the fulfillment of the strong field assumption at low magnetic fields has remained virtually unresolved. Perhaps, its certain decision is to use the pairs of edge channels in topological two-dimensional insulators and superconductors, in which the carriers with anti-parallel spins move in opposite directions [7]. Especially as recently the ideas are suggested that the mobility and spin-lattice relaxation time of the carriers in topological channels can be increased, if to hide them in the cover consisting of the $d$- and $f$-like impurity centers [8,9]. Here we use as similar clothes the striations of the negative-$U$ dipole boron centers that allow except noted advantages to achieve the effective cooling inside the edge channels of semiconductor quantum wells in the process of the spin-dependent transport. The aforesaid promotes also the creation of composite bosons and fermions by the capture of single magnetic flux quanta on the edge channels under the conditions of low sheet density of carriers, thus opening new opportunities for the registration of the quantum kinetic phenomena in weak magnetic fields at high temperatures up to the room temperature. As a certain version noted above we present the first findings of the high temperature de Haas-van Alphen, 300 K, and quantum Hall, 77 K, effects as well as the quantum conductance staircase in the silicon sandwich structure that represents the ultra-narrow, 2 nm, p-type quantum well (Si-QW) confined by the delta barriers heavily doped with boron on the *n*-type Si (100) surface. These data appear to result from the low density of single holes that are of small effective mass in the edge channels of *p*-type Si-QW because of the impurity confinement by the stripes consisting of the negative-*U* dipole boron centers which seems to give rise to the efficiency reduction of the electron-electron interaction.

## 2. Experiment

The device was prepared using silicon planar technology. After precise oxidation of the *n*-type Si (100) wafer, making a mask and performing photolithography, we have applied short-time low-temperature diffusion of boron from a gas phase [10,11]. Finally, the ultra-shallow $p^+$–$n$ junction has been identified, with the $p^+$ diffusion profile depth of 8 nm and the extremely high concentration of boron, $5 \cdot 10^{21}$ cm$^{-3}$, according to the SIMS data (Fig. 1(a)) [10]. Next step was to apply the cyclotron resonance (Fig. 1(b)), the electron spin resonance, the tunneling spectroscopy (Fig. 1(c)), infrared Fourier spectroscopy methods as well as the measurements of the quantum conductance staircase for the studies of the quantum properties of the silicon nanosandwich structures [11–14].

Firstly, the cyclotron resonance angular dependences have shown that the $p^+$ diffusion profile contains the ultra-narrow p-type silicon quantum well, Si-QW, confined by the wide-gap delta-barriers heavily doped with boron (Fig. 1(a)) [12,13]. Secondly, the one-electron band scheme for the delta barriers and the energy positions of two-dimensional subbands of holes in the Si-QW have been revealed using the tunneling and the infrared Fourier spectroscopy techniques (Fig. 1(d)) [11,13]. Then, the studies of the spin interference by measuring the Aharonov–Casher oscillations allowed the identification of the extremely low value of the effective mass of holes [11]. These data have been also confirmed by measuring the temperature dependence of the ShdH and dHvA oscillations [13,15].

The planar silicon sandwich structures prepared were very surprised to demonstrate the high mobility of holes in the Si-QW in spite of the extremely high concentration of boron inside the delta barriers. Specifically, the cyclotron resonance spectra exhibit the long moment relaxation time for both heavy and light holes, $>5 \cdot 10^{-10}$ s, and electrons, $>2 \cdot 10^{-10}$ s (Fig. 1(b)) [12,13]. These results appeared to be caused by the formation of the trigonal dipole boron centers, $B^+$–$B^-$, due to the negative-*U* reaction: $2B^0 \rightarrow B^+ + B^-$ (Fig. 1(e)) [11,14]. The excited triplet states of the negative-*U* centers were observed firstly by measuring the electron spin resonance angular dependences. It is important that the ESR spectra of the negative-*U* centers are revealed only after cooling in magnetic fields above critical value of 0.22 T, with persistent behavior in the dependence of temperature variations and crystallographic orientations of magnetic field under cooling procedure [16]. Such persistent behavior of the ESR spectra seems to be evidence of the dynamic magnetic moment due to the arrays of the trigonal dipole boron centers which dominate inside the delta-barriers confining the *p*-type Si-QW [11,13,16].

The scanning tunneling microscopy (STM), as well as the scanning tunneling spectroscopy, STS, studies showed that the negative-*U* dipole boron centers are able to form the stripes crystallographically oriented along the [110] axes (Figs. 1(f) and 1(g)) [11]. Moreover, the subsequences of these stripes are able to create the edge channels that define the conductance of the silicon nanosandwiches. And, if the sheet density of holes is rather small, on each such stripe no more than one hole settles down that appears to result in the neutralization of the electron-electron interaction owing to the exchange interaction with the negative-*U* dipole boron centers. Therefore we could apparently control the step-by-step capture of single magnetic flux quanta on the negative-*U* dipole boron stripes containing single holes by measuring the field dependences of the static magnetic susceptibility. We have carried out these experiments with the long edge channel, 2 mm, of the *p*-type Si-QW, taking account of the sheet density of holes, $3 \cdot 10^{13}$ m$^{-2}$, measured with similar device prepared within frameworks of the Hall geometry (Fig. 1(h)). This value of the sheet density of holes specified previously that each stripe seems to contain a single hole in this edge channel (Fig. 1(i)).

### 2.1. Static magnetic susceptibility measurements.

Figures 2(a) and (b) show the room temperature field dependences of the static magnetic susceptibility of the silicon nanosandwiches that have been obtained by the Faraday method. The high sensitivity, $10^{-9}$–$10^{-10}$ GGS, of the balance spectrometer MGD31FG provided the high stability that is necessary to calibrate the $BdB/dx$ values. In turn, for the $BdB/dx$ calibration, we used pure InP single crystals that are similar to investigated samples in shape and size, and reveal the high temperature stability of the magnetic susceptibility value, $\chi = -313 \cdot 10^{-9}$ cm$^3$/g [13,15].

Owing to so high sensitivity of the balance spectrometer up to weak magnetic fields, the oscillations of the static magnetic susceptibility that is due to the step-by-step capture of single magnetic flux quanta appeared to be revealed against the background of the strong diamagnetism of the negative-*U* dipole boron stripes (Figs. 2(c)) and (d)). The half-circle value of these oscillations is evidence of such an assumption, with the alternation of the spin-up and spin-down for holes under these step-by-step processes in the edge channel, if the size of the edge channel is taken into account, $\Phi_0 = \Delta\Phi = \Delta BS$, where $\Delta B = 10$ G; $S = 2$ mm × 2 nm = $= 4\cdot10^{-12}$ m$^2$ is the area of the edge channel; $\Phi_0 = h/e$ is the single magnetic flux quantum. Thus, the step-by-step change of a magnetic field that is equal to $\Delta B = 10$ G is able to result in the capture of single magnetic flux quanta on each hole after 124 circles, which correspond approximately to the number of them in the edge channel in view of the sheet density value. Under these conditions, $\Phi_0 = \Delta BS = B_0 S_0$, where $B_0 = 124·\Delta B = 1240$ G, $S_0$ is the area occupied by a single hole inside the negative-$U$ dipole boron stripes which is defined by the middle distance between holes in the edge channel, $S_0 = S/124 = 16.6$ microns. So, there is a unique possibility to create composite bosons since weak magnetic fields that result from the capture of single magnetic flux quanta on the holes which are poorly interacting with each other in the presence of the negative-$U$ dipole boron stripes.

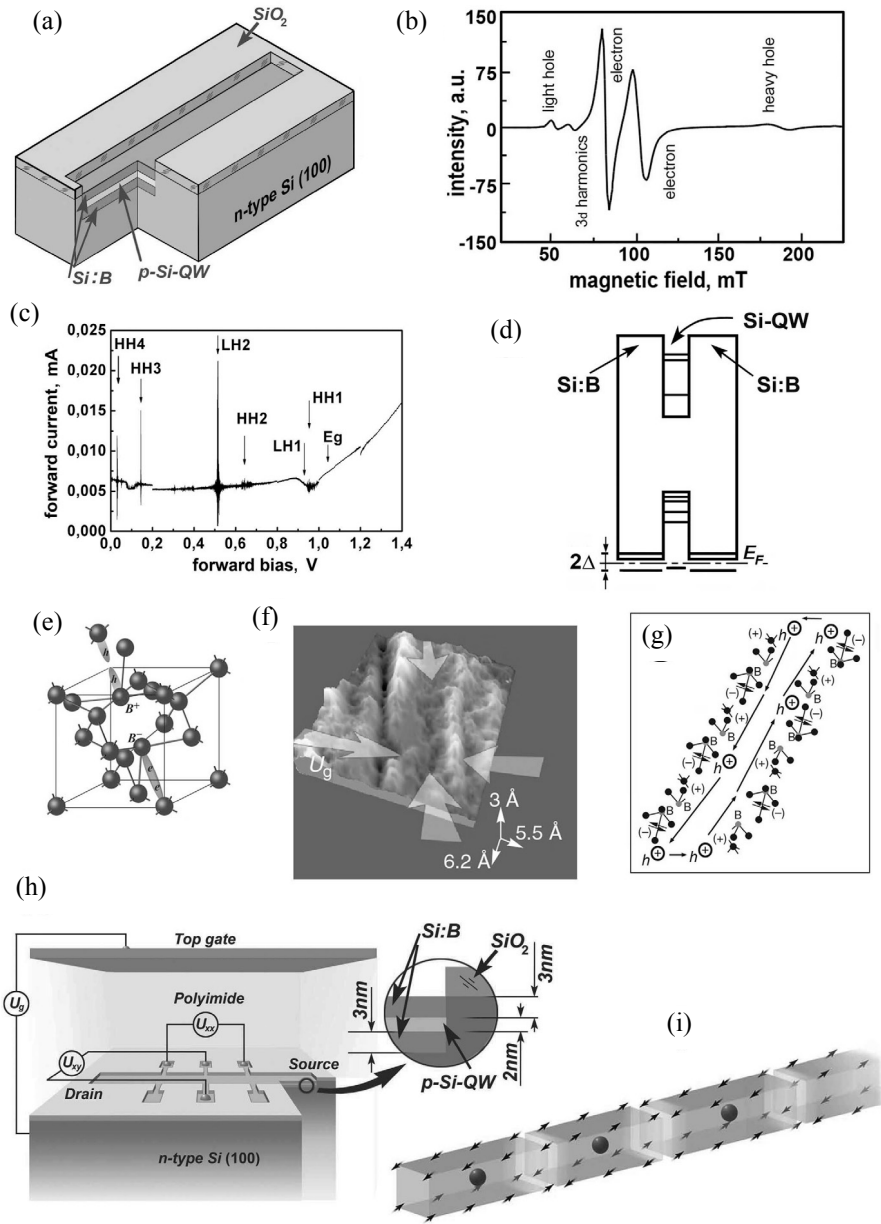

*Fig. 1.* Planar silicon nanosandwiches. (a) Scheme of the silicon nanosandwich that represents the ultra-narrow *p*-type silicon quantum well (Si-QW) confined by the delta barriers heavily doped with boron on the *n*-type Si (100) wafer. (b) Cyclotron resonance (CR) spectrum for a QW $p^+$–$n$ junction on {100}-silicon (500 Ohm·cm), magnetic field *B* within the {110}-plane perpendicular to the {100}-interface ($B_\parallel < 100 > + 30°$); *T* = 4.0 K, $\nu$ = 9.45 GHz. (c) The current-voltage characteristics under forward bias applied to the silicon nanosandwich. The energy position of each subband of 2D holes in Si-QW is revealed as a current peak under optimal tunneling conditions when it coincides with Fermi level. *T* = 300 K. (d) The one-electron band scheme of the *p*-type Si-QW confined by the delta-barriers heavily doped with boron on the *n*-type Si (100) surface. $2\Delta$ depicts the correlation gap in the delta barriers that results from the formation of the negative-*U* dipole boron centers. (e) Model for the elastic reconstruction of a shallow boron acceptor which is accompanied by the formation of the trigonal dipole ($B^+$–$B^-$) centers as a result of the negative-*U* reaction: $2B^0 \rightarrow B^+ + B^-$. (f) STM image of the upper delta barrier heavily doped with boron that demonstrates the chains of dipole boron centers oriented along the [011] axis. (g) The fragments of the edge channels that contain the stripes consisting of the negative-*U* dipole boron centers. (h) Experimental device prepared within the Hall geometry based on an ultra-narrow *p*-type silicon quantum well (Si-QW) confined by the delta barriers heavily doped with boron on the *n*-Si (100) surface. (i) The model of the edge channel in the silicon nanosandwich that is confined by the impurity stripes containing single holes.

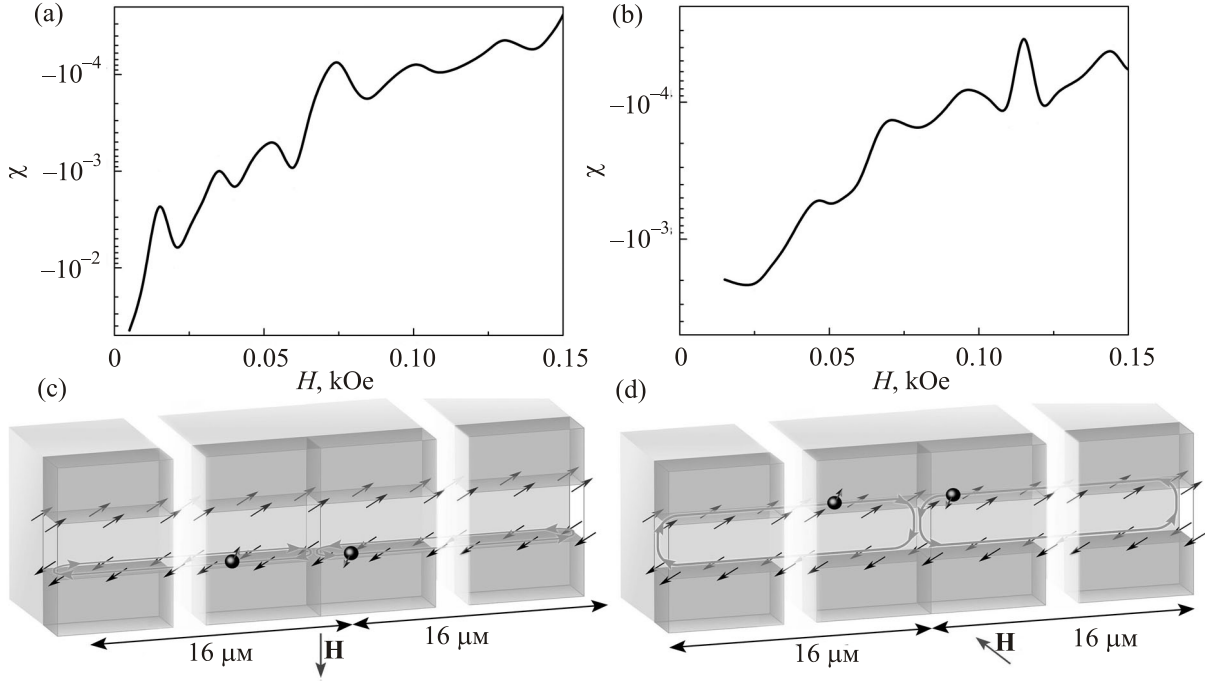

*Fig. 2.* Step-by-step capture of magnetic flux quanta on the edge channel in silicon nanosandwich. (a) and (b) The field dependences of the static magnetic susceptibility at room temperature in silicon nanosandwich, $p_{2D} = 3·10^{13}$ m$^{-2}$, which reveal strong diamagnetism that is caused by the spin precession of single holes confined by the negative-$U$ dipole boron stripes in the edge channel and the oscillations due to the step-by-step capture of single magnetic flux quantum on the edge channel at perpendicular (a) and (c) and parallel (b) and (d) orientation of a magnetic field related to the Si-QW plane. $T = 300$ K.

These reasons are supported by the analysis of huge amplitudes of the static magnetic susceptibility oscillations which appears to give rise to the values of magnetization, $J = \chi H$, up to 0.2 Oe. From here it is possible to estimate the value of the magnetic moment created by the capture of the magnetic flux quantum on a single hole in view its volume, $M = JV_0$, where $V_0 = LS_1$; $L$ is the middle distance between holes in the edge channel and $S_1$ is the cross section of the edge channel. This assessment results in the value of the magnetic moment, $2.4·10^4$ $m_B$, that unambiguously points to a fundamental role of the stripes consisting of the negative-$U$ dipole boron centers in its formation, where $m_B$ is the Bohr magneton. And really, if we estimate the number of these centers in the volume occupied by a single hole in view of the concentration of boron, $5·10^{21}$ cm$^{-3}$, and having compared everyone the magnetic moment equal to the Bohr magneton, it is possible to be convinced that it practically coincides with the value given above. Similar results are important to be obtained in the case of both parallel and perpendicular orientation of the magnetic field to the Si-QW plane, because the edge channel has square section (see Figs. 2(c)) and (d)). It is also necessary to note supervision of almost limit value of a diamagnetic static susceptibility in weak magnetic fields, $\chi = 1/4\pi$, that appears to be due to superconducting properties of edge channels because of the formation of the negative-$U$ dipole boron centers.

Thus, the magnetic susceptibility response to the capture of magnetic flux quanta on the edge channel is caused by the magnetic ordering of the stripes through single holes. This exchange interaction seems to lead to partial localization of single holes and as a result to the reduction of electron-electron interaction. Moreover, by increasing the magnetic field the static magnetic susceptibility begins to reveal the dHvA oscillations due to the creation of the Landau levels, $E_\nu = \hbar\omega_c(\nu + ½)$, $\nu$ is the number of the Landau level. In particular, a prerequisite to cover the edge channel by single magnetic flux quanta appears to be accomplished when the external magnetic field is equal to $B_0 = 1240$ G, see the relationship presented above, $\Phi_0 = \Delta B·S = B_0·S_0$, that is consistent with the first Landau level feeling, $\nu_1 = 1$, $\nu_1 = p_{2D}h/eB_0$.

We studied carefully the dHvA oscillations created at both parallel and perpendicular orientation of the magnetic field to the Si-QW plane and registered except the dips related to the Landau levels, $\nu = 1, 2, 3, 4, 5, 6$, the fractional peaks, $\nu = 4/3$ and $\nu = 5/3$ (Figs. 3(a)–(f)). These findings demonstrate that in certain ratio between the number of magnetic flux quanta and single holes, $1/\nu$, both composite bosons and fermions seem to be proceeded by step-by-step change of a magnetic field. Besides, using specific diagram of magnetic field–edge channel feeling, the variations of integer and fractional values of $\nu$ illustrate conveniently their relationship by varying the external magnetic field (Fig. 4). It should be noted also that the strong diamagnetism of the negative-$U$ impurity stripes surrounding the Si-QW edge channels allowed the observation of the room temperature hysteresis of the dHvA oscillations dependent on the proximity of the Landau and Fermi levels. In turn, the creation of the composite bosons ever in weak magnetic fields inside stripes containing single holes can lead to the emergence of the Faraday effect under the conditions of the source - drain current in the edge channel, $I_{ds} = dE/d\Phi$. This model has been suggested by Laughlin to account for the quantum staircase of the Hall resistance [17]. Here we present the results of the measurements of not only integer but also fractional quantum Hall effect in the same magnetic field as well as the dHvA oscillations, $I_{ds} = I_{xx} = eU_{xy}/(1/\nu)\Phi_0, \Rightarrow G_{xy} = \nu e^2/h$, where $\nu$ can accept both integer and fractional values.

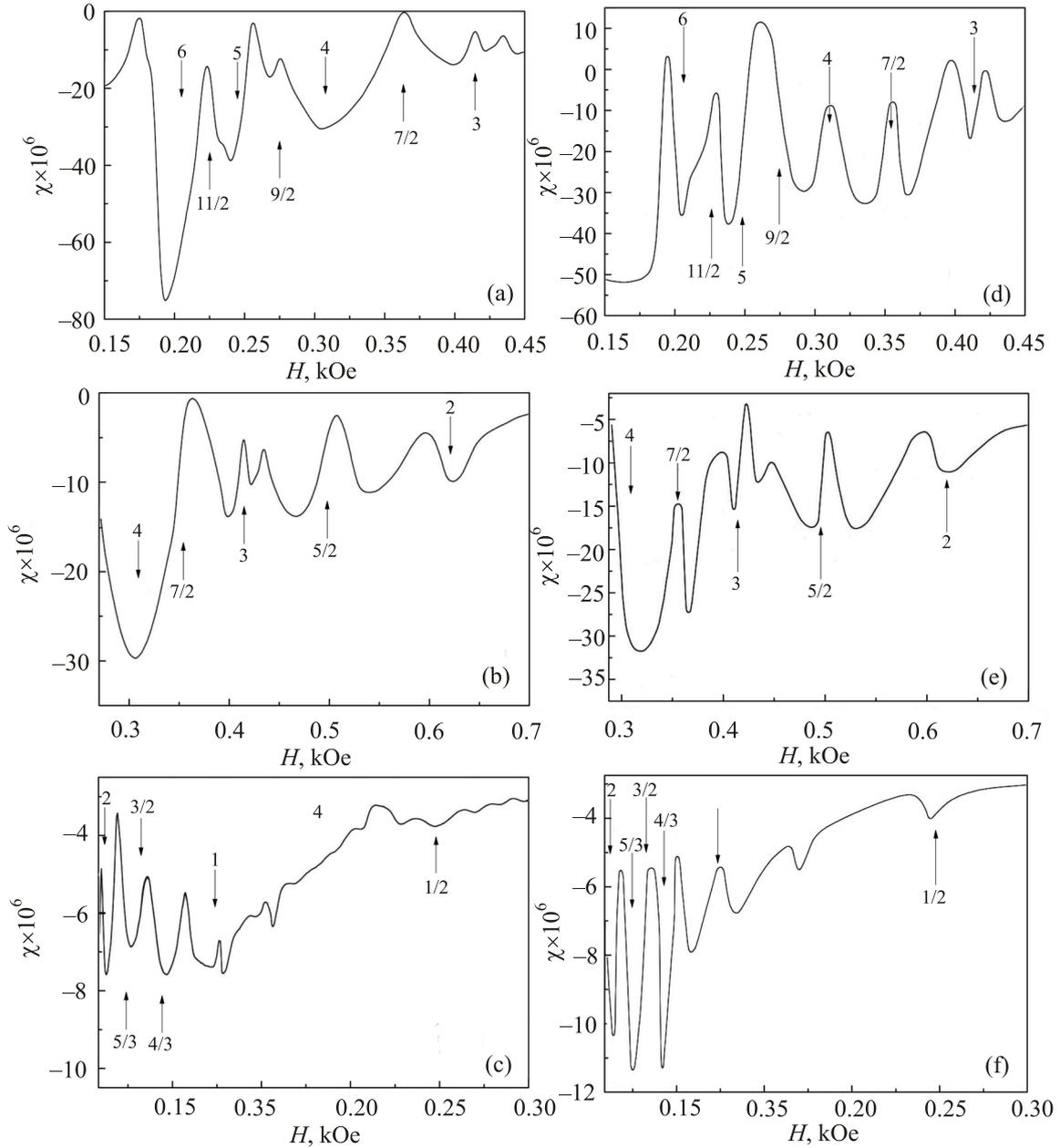

*Fig. 3.* The de Haas–van Alphen oscillations. The de Haas–van Alphen oscillations revealed in the field dependences of the static magnetic susceptibility measured at room temperature in silicon sandwich, $p_{2D} = 3 \cdot 10^{13}$ m$^{-2}$, at perpendicular (a), (b), (c) and parallel (d), (e), (f) orientation of a magnetic field related to the Si-QW plane. The dips is related to the Landau levels $\nu = 1, 2, 3, 4, 5, 6$. $T = 300$ K.

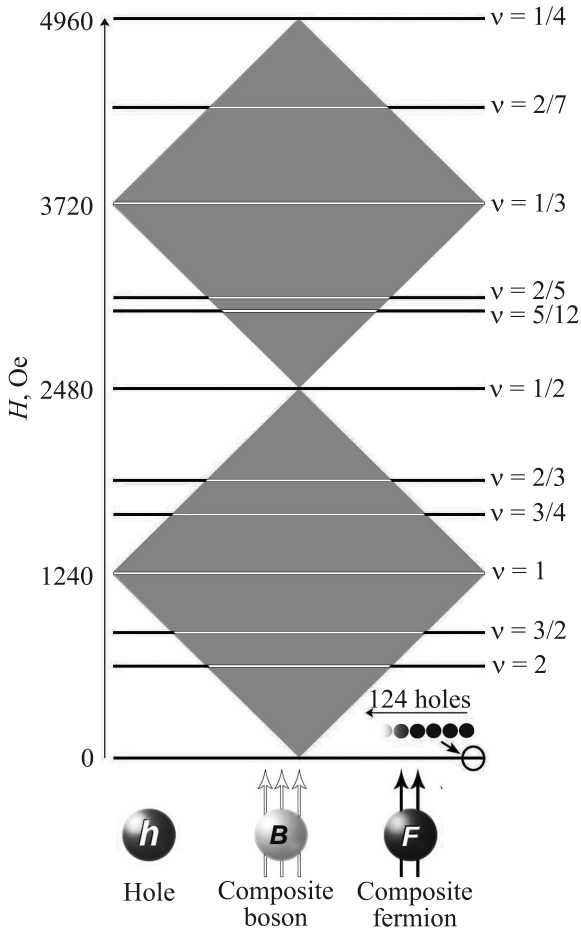

*Fig. 4.* Composite bosons and fermions. Diagram showing the subsequent creation of composite bosons (light line) and fermions (dark line) that result from the capture of single magnetic flux quantum at the edge channel of silicon nanosandwich confined by the negative-*U* impurity stripes containing single holes.

### 2.2. Quantum Hall effect

Firstly, the SdH oscillations and the quantum Hall staircase are demonstrated, with the identification of both the integral and fractional quantum Hall effects (Figs. 5(a), (b) and (c)). Secondly, the range of a magnetic field corresponding to the Hall plateaus and the longitudinal "zero" resistance is in a good agreement with the interval of the dHvA oscillations thereby verifying the principal role of the Faraday effect in these processes. Here, the confinement of single holes inside the stripes consisting of the negative-*U* dipole boron centers seems to lead not only to the efficiency reduction of the electron-electron interaction, but also promotes quantization of the interelectronic spacing [18–21]. Thus, the stabilization of a ratio between the number of magnetic flux quanta and single holes in edge channels, $1/\nu$, is reached at certain values of an external magnetic field, thereby promoting the registration of both integer, and fractional quantum Hall effect [22,23].

In addition to the aforesaid, the DX- and oxygen-related centers as well as the antisite donor–acceptor pairs reveal the negative-*U* properties to confine effectively the edge channels in III–V compound low-dimensional structures [24–27]. However the preparation of dipolar configurations in this case appears to be accompanied by preliminary selective illumination at low temperatures.

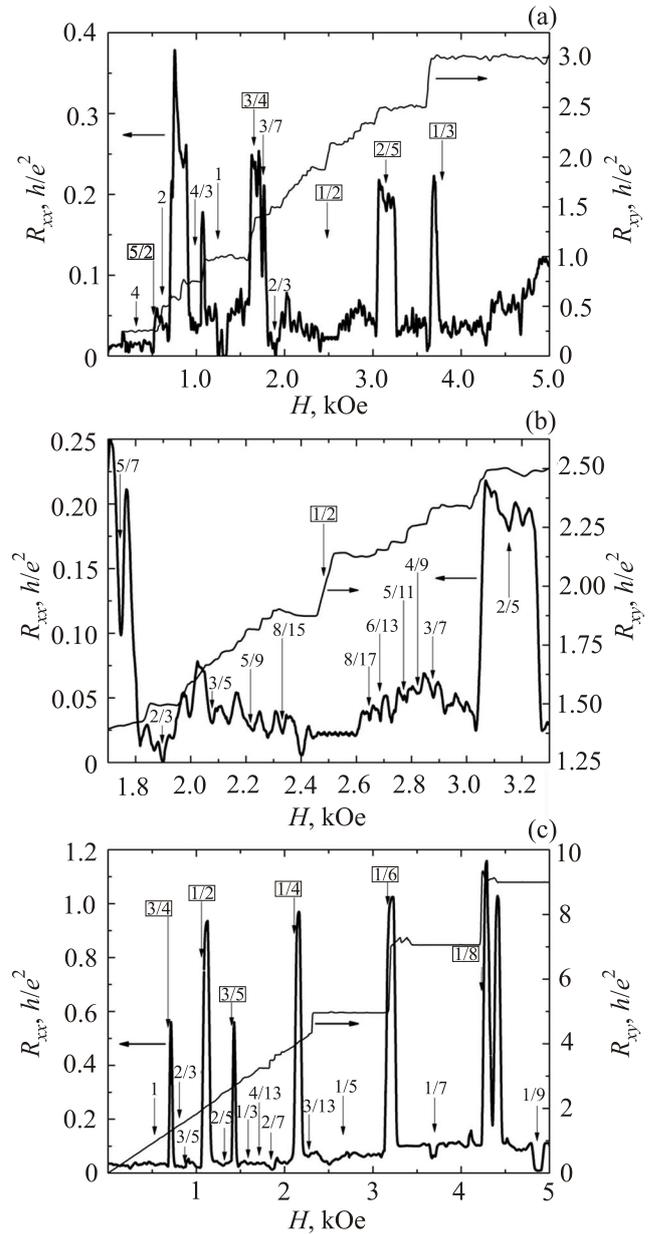

*Fig. 5.* Quantum Hall effect. (a) The Hall resistance $R_{xy} = V_{xy}/I_{ds}$ and the magnetoresistance $R_{xx} = V_{xx}/I_{ds}$, $I_{ds} = 10$ nA, of a two-dimensional hole system, $p_{2D} = 3 \cdot 10^{13}$ m$^{-2}$, in silicon nanosandwich at the temperature of 77 K vs magnetic field. (b) Manifestation of the fractional quantum Hall effect near $\nu = 1/2$, $p_{2D} = 3 \cdot 10^{13}$ m$^{-2}$. (c) $R_{xx}$ and $R_{xy}$ vs magnetic field at the temperature of 77 K, $I_{ds} = 10$ nA, for low sheet density of holes, $p_{2D} = 1.3 \cdot 10^{13}$ m$^{-2}$, that is revealed by varying the value of the top gate voltage (Fig. 1(h)), $V_{tg} = +150$ mV, see Ref. 11.

Nevertheless, the question arises — how is possible to measure these quantum effects at high temperatures in weak magnetic fields? One of the reasons that are noted above is small effective mass, $10^{-4}$ $m_0$, which is considered within the concept of the squeezed silicon arising owing to the negative-*U* impurity stripes in edge channels (Fig. 6) [11]. Other reason is caused by very effective self-cooling inside negative-*U* dipole boron strata under the conditions of the drain-source current.

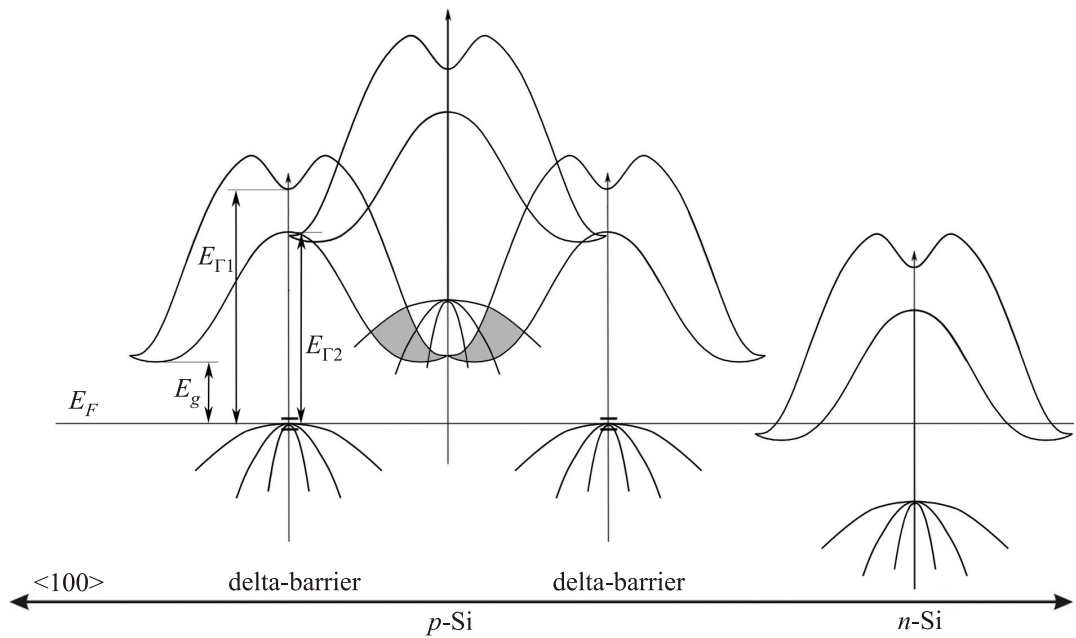

*Fig. 6.* One-electron band scheme of the silicon nanosandwich. The one-electron band scheme of the *p*-type Si-QW confined by the delta barriers heavily doped with boron on the *n*-type Si (100) surface. The delta-barriers are of wide-gap as a result of the formation of the negative-*U* dipole boron centers (see also Fig. 1(d)). The negative-*U* impurity stripes are of importance to result in the squeeze of the *p*-type Si-QW between the delta-barriers. In the frameworks of the squeezed silicon, the anti-crossing between the conduction band of the delta barriers and the valence band of the *p*-type Si-QW becomes to be actual thereby giving rise to the small effective mass of holes[11].

In order to interpret the experimental data presented above, we propose the following thermodynamic model of the self-cooling process in the silicon nanosandwich. In the frameworks of this model, the whole thermodynamic process consists of multiple self-cooling cycles, each of which includes the following two stages: A) the adiabatic electric depolarization process of the spontaneously polarized stripe that is due to the presence of holes in the edge channel, and B) the fast spontaneous isothermal polarization of the negative-$U$ boron dipoles in the stripe with transition back into the ferroelectric state.

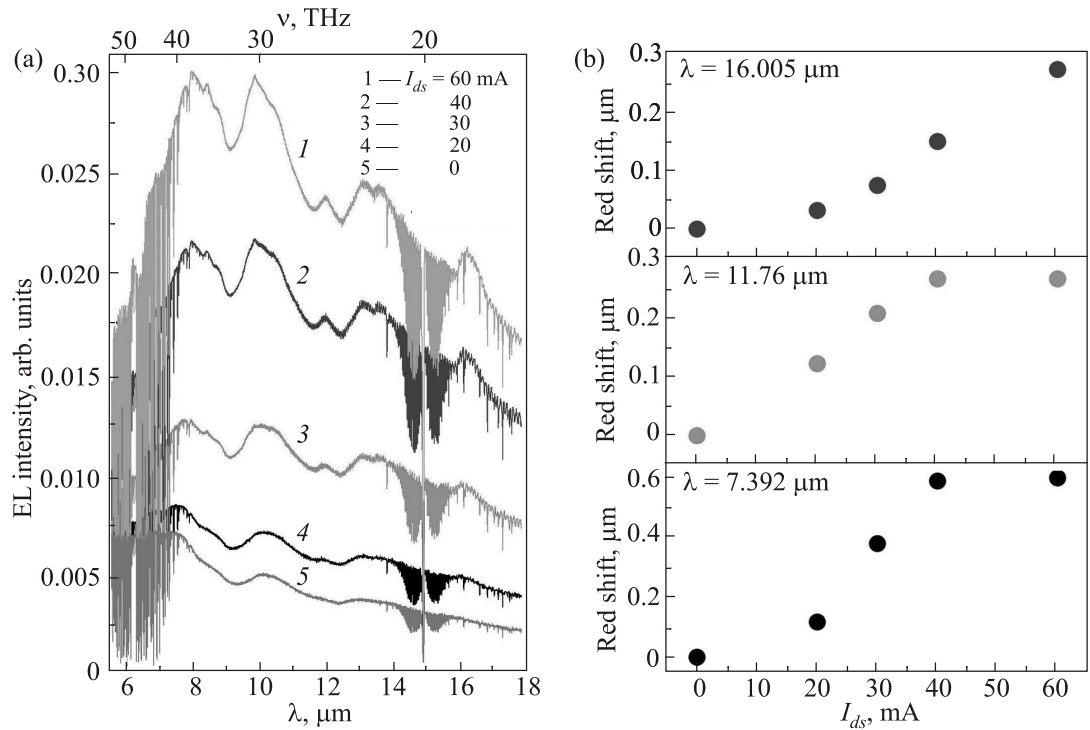

*Fig. 7.* The red shift of the radiation from the edge channels of silicon nanosandwich. (a) and (b) the cooler effect is identified by the red shift of the radiation caused by the dipole boron centers from the edge channels of the silicon nanosandwich device (see Fig. 1(h)) that is induced by the stabilized drain-source current in different spectral range; $T = 300$ K.

The model presented is based on 1) the energy conservation law in the electric depolarization process, and 2) the expression for the free energy, $f$, and the internal energy, $u$, of the ferroelectric state related to the thermodynamic work reserved in the edge channel of the silicon nanosandwich,

$$\delta Q = T\delta s = \delta u + \int E \delta P dV, \quad (1)$$

$$u = -T^2 \frac{\partial}{\partial T}\left(\frac{f}{T}\right) = \int T^2 \frac{\partial}{\partial T}\left(\frac{EP}{T}\right)dV. \quad (2)$$

Here $T$ and $s$ are the local effective temperature and the entropy of the nanosystem under consideration.

Combining the equations (1) and (2), we obtain the fundamental relationship describing a single thermodynamic self-cooling cycle of the negative-$U$ stripe that is induced by the drain-source current in the edge channel

$$c_p dT = -T\left(\frac{\partial E}{\partial T}\right)_p dP, \quad \Delta S = 0. \quad (3)$$

Here $E$, $P$ and $c_p$ are related to the local electric field, spontaneous dielectric polarization and the local lattice thermocapacity of the negative-$U$ stripe. Integration reveals a significant cooling effect in the edge channel of the silicon nanosandwich.

Thus, this model allows our interpretation of high temperature quantum kinetic effects in silicon nanosandwiches confined by the negative-$U$ ferroelectric or superconductor barriers. Specifically, the self-cooling effect of the edge channels can be derived as a red shift in the electroluminescence spectra that is enhanced by increasing the drain-source current, see Fig. 7.

### 2.3. Quantum conductance staircase in edge channels of silicon nanosandwiches

At present, the methods of a semiconductor nanotechnology such as the split-gate [28–30] and cleaved edge overgrowth [31] allow the fabrication of the quasi-one-dimensional (1D) constrictions with low density high mobility carriers, which exhibit the characteristics of ballistic transport. The conductance of such quantum wires with the length shorter than the mean free path is quantized in units of $G_0 = g_s e^2/h$ depending on the number of the occupied 1D channels, $N$. This quantum conductance staircase, $G =$ $G_0 N$, has been revealed by varying the split-gate voltage applied to the electron and hole GaAs [28–30] and Si-based [11] quantum wires. It is significant that spin factor, $g_s$, that describes the spin degeneration of carriers in a 1D channel appears to be equal to 2 for noninteracting fermions if the external magnetic field is absent and becomes unity as a result of the Zeeman splitting of a quantum conductance staircase in strong magnetic field.

However, the first step of the quantum conductance staircase has been found to split off near the value of $0.7(2e^2/h)$ in a zero magnetic field. Two experimental observations indicate the importance of the spin component for the behavior of this $0.7(2e^2/h)$ feature. Firstly, the electron $g$ factor was measured to increase from 0.4 to 1.3 as the number of occupied 1D subbands decreases [32]. Secondly, the height of the $0.7(2e^2/h)$ feature attains a value of $0.5(2e^2/h)$ with increasing external magnetic field [32]. These results have defined the spontaneous spin polarization of a 1D gas in a zero magnetic field as one of possible mechanisms for the $0.7(2e^2/h)$ feature in spite of the theoretical prediction of a ferromagnetic state instability in ideal 1D system in the absence of magnetic field [33].

More recently, the $0.7(2e^2/h)$ feature has been shown to be close also to the value of $0.5(2e^2/h)$ at low sheet density of both holes [11] and electrons [34]. These measurements performed by tuning the top gate voltage that allows the sheet density control have revealed not only the principal role of the spontaneous spin polarization, but the changes of the $0.7(2e^2/h)$ feature in fractional form as well [11]. Therefore the studies of the $0.7(2e^2/h)$ feature are challenging to be explored in the topological insulators or superconductors in which the one-dimensional, 1D, edge channels are formed without any electrical or mechanical restriction [35,36]. These edge channels are similarity to the quantum Hall states wherein the electric current is carried only along the edge of the sample, but exhibit helical properties, because two carriers with opposite spin polarization counter propagate at a given edge [37].

However, the quantum conductance staircase caused by the carriers in the edge channels is still uninvestigated in a zero magnetic field. Here we demonstrate the quantum conductance staircase of holes that is revealed by varying the voltage applied to the Hall contacts prepared at edges of the ultra-narrow p-type silicon quantum well, Si-QW. The fractional features that emerge in this conductance staircase appear to be evidence of the high spin polarization of holes in the helical edge channels. As noted above, the latter findings were made possible by the developments of the diffusion nanotechnology that allows the fabrication of the nanosandwiches prepared on the *n*-type Si (100) surface as the ultra-narrow p-type Si-QW, 2 nm, confined by the delta-barriers heavily doped with boron (see Fig. 1h) [11].

These characteristics of the 2D gas of holes in the silicon nanosandwiches made it possible for the first time to use the split-gate constriction to study the $0.7(2e^2/h)$ fea-ture in the quantum conductance staircase of holes at the temperature of 77 K [11–16]. Furthermore, even with small drain-source voltage the electrostatically ordered dipole centres of boron within the delta-barriers appeared to stabilize the formation of the one-dimensional subbands, when the quantum wires are created inside Si-QW using the split-gate technique [11]. Thus, the ultra-narrow p-type Si-QW confined by the delta-barriers heavily doped with boron in properties and composition seems to be similar to graphene [4,5] that enables one to observe the fractional quantum conductance staircase of holes by varying the voltage applied to the Hall contacts which is able to control the changes of the spin polarization in the edge channels (see Fig. 1(h)).

The $R_{xx}$ dependence on the bias voltage applied to the Hall contacts, $V_{xy}$, exhibits the quantum conductance staircase to a maximum of $4e^2/h$ (Fig. 8(a)). This conductance feature appeared to be independent of the sample geometric parameters that should point out on the formation of the edge channels in Si-QW. Therefore we assume that the maximum number of these channels is equal to 2, one for up-spin and other for down-spin. It should be noted that the important condition to register this quantum conductance staircase is to stabilize the drain-source current at the value of lower than 1 nA.

In addition to the standard plateau, $2e^2/h$, the quantum conductance staircase appears to reveal the distinguishing features as the plateaus and steps that bring into correlation respectively with the odd and even fractions. Since similar quantum conductance staircase was observed by varying the top gate voltage which controls the sheet density of carriers and thus can be favorable to the spontaneous spin polarization, the variations of the $V_{xy}$ value seem also to result in the same effect on the longitudinal resistance, $R_{xx}$ (Figs. 8(b) and (c)).

The $R_{xx}$ fractional values revealed by tuning the $V_{xy}$ voltage appear to evidence that the only closely adjacent helical channels to the edge of Si-QW make dominating contribution in the quantum conductance staircase as distinguished from the internal channels. The device used implies a vertical position of helical channels, with quantum point contact (QPC) inserted as a result of the local disorder in the delta-barriers heavily doped with boron. Besides, if the silicon nanosandwich is taken into account to be prepared along the [011] axis (Fig. 1(h)), the trigonal dipole boron centers ordered similarly appear to give rise to the formation of the helical edge channels in Si-QW. It should be noted that depending on the degree of disorder in the delta-barriers the helical channels have to reveal the insulating or superconducting properties thereby defining the regime of the spin-dependent transport through the ballistic QPC [38]. In the latter case the multiple Andreev reflections seem to result in the spin polarization of carriers in helical channels in addition to the mechanism of spontaneous spin polarization [13].

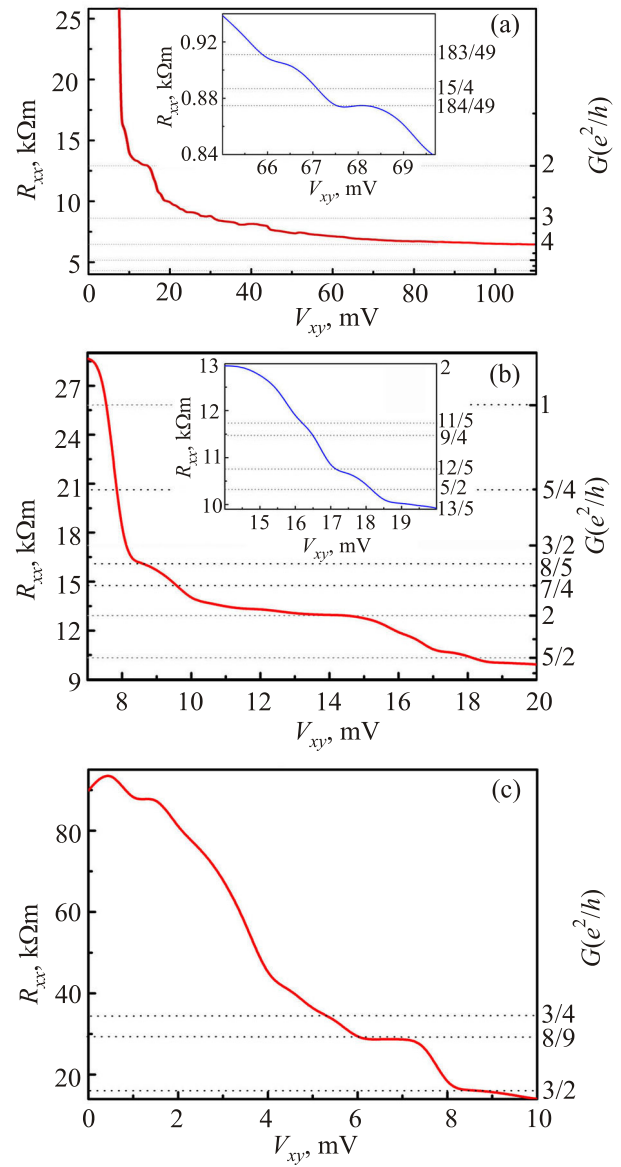

*Fig. 8.* Conductance measured at the temperature of 77 K by biasing the voltage applied to the Hall contacts, $V_{xy}$, when the drain-source current was stabilized at the value of 0.5 nA. (a) Conductance increases as a function of $V_{xy}$ to a maximum of $4e^2/h$ demonstrating the standard plateaus, $2e^2/h$ and $3e^2/h$. The inset shows the conductance feature at the value of $15/4(e^2/h)$. (b) Fractional quantum conductance staircase close to the standard plateau at the value of $2e^2/h$. Insert shows the conductance plateaus and steps corresponding to the odd and even fractions. (c) Conductance measured as a function of $V_{xy}$ in the range of the values corresponding to the $0.7(2e^2/h)$ feature.

The exchange interaction between holes localized and propagating through the QPC inside a quantum wire has been shown to give rise to the fractional quantization of the conductance unlike that in electronic systems [39]. Both the offset between the bands of the heavy and light holes, $\Delta$, and the sign of the exchange interaction constant appeared to affect on the observed value of the conductance at the additional plateaus. Within the framework of this approach for the Si-based quantum wire, the conductance plateaus have to be close to the values of $e^2/4h$, $e^2/h$ and $9e^2/4h$ from the predominant antiferromagnetic interaction, whereas the prevailing ferromagnetic exchange interaction has to result in the plateaus at the values of $7e^2/4h$, $3e^2/h$ and $15e^2/4h$ [39]. Bearing in mind this prediction, we have observed the fractional features, which are in a good agreement with the ferromagnetic interaction theory (Figs. 2(a), (b) and (c)). But the reason why the even fractional values correspond to the middle of the steps in the quantum conductance staircase instead of the plateaus as predicted in Ref. 39 is needed to be analyzed in detail. Since the odd and even fractions have been exhibited similarly in the quantum conductance staircase as a function of the top gate voltage that controls the sheet density of holes, the parallels between the helical channels and the ballistic channels responsible for the fractional quantum Hall effect in strong magnetic fields engage once again our attention. As mentioned above, the helical edge channels could be used to verify the interplay between the amplitude of the $0.7(2e^2/h)$ feature and the degree of the spontaneous spin polarization when the energy of the ferromagnetic exchange interaction begins to exceed the kinetic energy in a zero magnetic field. Specifically, the evolution of the $0.7(2e^2/h)$ feature in the quantum conductance staircase from $e^2/h$ to $3/2(e^2/h)$ is shown in Fig. 2(c), which seems to be caused by the spin depolarization processes in QPC [11,14,39]. It is appropriate to suggest that the spin depolarization is the basic mechanism of the quantum conductance staircase as a function of the $V_{xy}$ bias voltage. This consideration can be best be done within framework of the Landauer–Buttiker formalism [37], if the $V_{xy}$ bias voltage is used to backscatter helical edge channels only on the one side of the device. As a result of this backscattering, that is different in adjacent channels, in which the carriers counter propagate, the total spin polarization has to be varied thus providing the quantum conductance staircase as a function of the $V_{xy}$ bias voltage. Experimentally, the high resolved variations of the $V_{xy}$ bias voltage appeared to cause the observation of the exotic plateaus and steps (see the inset in Fig. 8(a)) that are evidence of the hole particles with fractional statistics [40], which are also needed to be studied in detail.

Nevertheless, within frameworks of the model suggested by Laughlin the quantum conductance in the edge channels appears to result from the magnetic moment of the stripes, which is induced by the stabilized drain-source current, $I_{ds}$. Here, the value of the longitudinal conductance is defined by the number of induced fluxes, $n_{ind}$, that are captured at stripes containing the single holes which can be analyzed within frameworks of the model of quantum harmonic oscillyator:

$$G_{xx} \sim \frac{e}{\partial \Phi_{ind}} = \frac{e}{n_{ind}\Phi_0} = \frac{1}{n_{ind}}\frac{e^2}{h}. \quad (4)$$

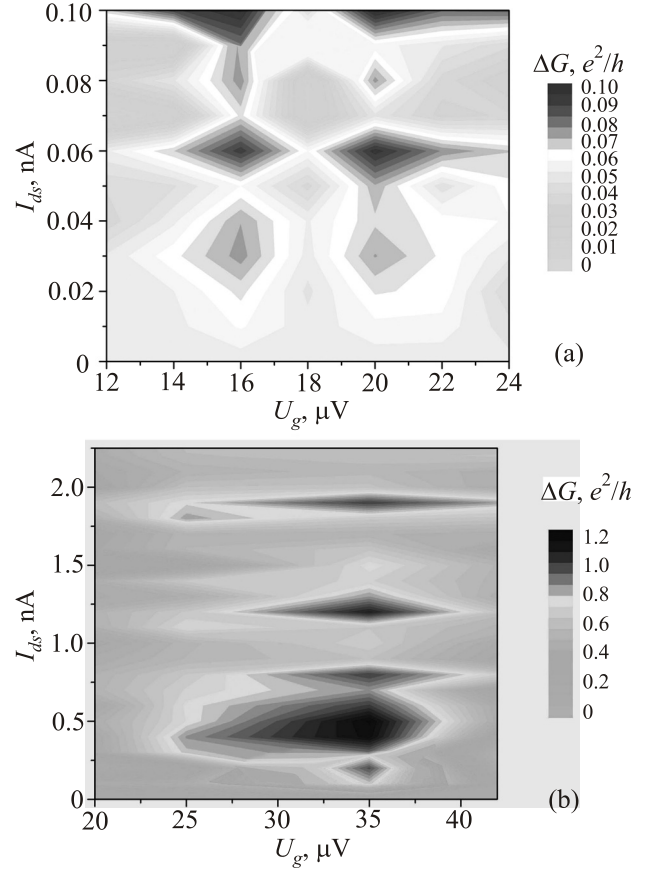

*Fig. 9.* Diagrams of the longitudinal conductance of holes at zero magnetic fields in a silicon sandwich device based on a *p*-type Si-QW confined by the delta-barriers heavily doped with boron on the *n*-Si (100) surface. The conductance is mapped as a function of gate voltage, which applied to the lateral contacts. (a) The minimum energy of the quantum harmonic oscillator formed by a single hole inside a dipole boron stripe is demonstrated; (b) fractal origin of the quantum interference in the topological channel in silicon sandwich device. *T* = 300 K.

Therefore the fractional values of the longitudinal conductance appear to be caused by the combinations of composite fermions and bosons that are possible to be created in both small and large values of the stabilized drain-source current (see. Figs. 9(a) and (b)). Besides, by varying the voltage applied to the Hall contacts, the phase coherence of spin-dependent transport in the edge channels seems to be controlled.

## Summary

Negative-*U* impurity stripes containing single holes appear to reduce the electron-electron interaction in the edge channels of the silicon nanosandwiches that allow the observations of quantum kinetic effects at high temperatures up to the room temperature.

The edge channels of semiconductor quantum wells confined by the subsequence of negative-*U* impurity stripes containing single holes give rise to the creation of the composite bosons and fermions in weak magnetic fields by the step-by-step capture of magnetic flux quanta.

The strong diamagnetism of the impurity stripes that consist of the negative-$U$ dipole boron centers surrounding the Si-QW edge channels allowed the high temperature observation of the de Haas–van Alphen oscillations as well as integer and the fractional quantum Hall effect.

We have also found the fractional form of the longitudinal quantum conductance staircase of holes, $G_{xx}$, that was measured as a function of the $V_{xy}$ bias voltage applied to the silicon nanosandwich prepared in the framework of the Hall geometry. The fractional conductance features observed at the values of $7e^2/4h$, $3e^2/h$ and $15e^2/4h$ are evidence of the prevailing ferromagnetic exchange interaction that gives rise to the spin polarization of holes in a zero magnetic field. This quantum conductance staircase measured to a maximum of $4e^2/h$, with the plateaus and steps that bring into correlation respectively with the odd and even fractional values, seems to reveal the formation of the helical edge channels in the p-type silicon quantum well.

Finally, it should be noted that the step-by-step capture of magnetic flux quanta on all negative $U$-impurity stripes in the edge channels is determined by the balance between the numbers of composite bosons and fermions which is also revealed by measuring the quantum conductance value, with controlled phase coherence by varying the stabilized drain-source current value even in the absence of the external magnetic field that seems to be very interesting for models of quantum computing.

The work was supported by the programme «5-100-2020», project 6.1.1 of SPSPU (2014); project 1963 of SPbPU (2014); the programme of fundamental studies of the Presidium of the Russian Academy of Sciences "Actual problems of low temperature physics" (grant 10.4); project 10.17 "Interatomic and molecular interactions in gases and condensed matter".